\documentclass[twocolumn,english,journal]{IEEEtran}
\usepackage[T1]{fontenc}
\usepackage{babel}
\usepackage{amsmath}
\usepackage{amssymb}
\usepackage[unicode=true,
 bookmarks=true,bookmarksnumbered=true,bookmarksopen=true,bookmarksopenlevel=1,
 breaklinks=false,pdfborder={0 0 0},backref=false,colorlinks=false]
 {hyperref}
\hypersetup{pdftitle={Your Title},
 pdfauthor={Your Name},
 pdfpagelayout=OneColumn, pdfnewwindow=true, pdfstartview=XYZ, plainpages=false}

\makeatletter
\usepackage[caption=false,font=footnotesize]{subfig}

\newtheorem{theorem}{Theorem}
\newtheorem{remark}{Remark}
\newtheorem{definition}{Definition}

\makeatother

\begin{document}

\title{Control of nonlinear systems:\\
a model inversion approach}

\author{C. Novara, M. Milanese%
\thanks{Carlo Novara is with Dipartimento di Automatica e Informatica, Politecnico
di Torino, Italy, e-mail: \protect\href{http://carlo.novara@polito.it}{carlo.novara@polito.it}.
Mario Milanese is with Modelway srl, Italy, e-mail: \protect\href{http://mario.milanese@modelway.it}{mario.milanese@modelway.it}.%
}}
\maketitle
\begin{abstract}
A novel control design approach for general nonlinear systems is presented
in this paper. The approach is based on the identification of a polynomial
model of the system to control and on the on-line inversion of this
model. An efficient technique is developed to perform the inversion,
which allows an effective control implementation on real-time processors.
\end{abstract}

\section{Introduction}

\label{sec:ibc_approach}

Consider a nonlinear discrete-time SISO system in regression form:
\begin{equation}
y_{t+1}=g\left(\boldsymbol{y}_{t},\boldsymbol{u}_{t},\boldsymbol{\xi}_{t}\right)\label{ss_sys}
\end{equation}
\[
\begin{array}[t]{c}
\boldsymbol{y}_{t}=\left(y_{t},\ldots,y_{t-n+1}\right)\\
\boldsymbol{u}_{t}=\left(u_{t},\ldots,u_{t-n+1}\right)\\
\boldsymbol{\xi}_{t}=\left(\xi_{t},\ldots,\xi_{t-n+1}\right)
\end{array}
\]
where $u_{t}\in U\subset\mathbb{R}$ is the input, $y_{t}\in\mathbb{R}$
is the output,$\xi_{t}\in\Xi\subset\mathbb{R}^{n_{\xi}}$ is a disturbance
including both process and measurement noises, and $n$ is the system
order. $U$ and $\Xi$ are compact sets. In particular, $U\doteq[\underline{u},\overline{u}]$
accounts for input saturation.

Suppose that the system (\ref{ss_sys}) is unknown, but a set of measurements
is available:
\begin{equation}
\mathcal{D}\doteq\left\{ \tilde{u}_{t},\tilde{y}_{t}\right\} _{t=1-L}^{0}\label{eq:data}
\end{equation}
where $\tilde{u}_{t}$ and $\tilde{y}_{t}$ are bounded for all $t=1-L,\ldots,0$.
The tilde is used to indicate the input and output samples of the
data set \eqref{eq:data}.

Let $\mathcal{Y}^{0}\subseteq\mathbb{R}^{n}$ be a set of initial
conditions of interest for the system (\ref{ss_sys}) and, for a given
initial condition $\boldsymbol{y}_{0}\in\mathcal{Y}^{0}$, let $\mathcal{Y}\left(\boldsymbol{y}_{0}\right)\subseteq\ell_{\infty}$
be a set of output sequences of interest.

The aim is to control the system (\ref{ss_sys}) in such a way that,
starting from any initial condition $\boldsymbol{y}_{0}\in\mathcal{Y}^{0}$,
the system output sequence $\boldsymbol{y}=(y_{1},y_{2},\ldots)$
tracks any reference sequence $\boldsymbol{r}=(r_{1},r_{2},\ldots)\in\mathcal{Y}\left(\boldsymbol{y}_{0}\right)$.
The set of all solutions of interest is defined as $\mathcal{Y}\doteq\left\{ \mathcal{Y}\left(\boldsymbol{y}_{0}\right):\boldsymbol{y}_{0}\in\mathcal{Y}^{0}\right\} $.
The set of all possible disturbance sequences is defined as $\varXi\doteq\left\{ \boldsymbol{\xi}=(\xi_{1},\xi_{2},\ldots):\xi_{t}\in\Xi,\forall t\right\} $.

After a brief section regarding the notation used in the paper, an
approach called NIC (Nonlinear Inversion Control) is proposed for
the design of a controller $K^{nl}$, allowing the accomplishment
of the above task.

\section{Notation}

A column vector $x\in\mathbb{R}^{n_{x}\times1}$ is denoted as $x=\left(x_{1},\ldots,x_{n_{x}}\right)$.
A row vector $x\in\mathbb{R}^{1\times n_{x}}$ is denoted as $x=\left[x_{1},\ldots,x_{n_{x}}\right]=\left(x_{1},\ldots,x_{n_{x}}\right)^{\top}$,
where $\top$ indicates the transpose.

A discrete-time signal (i.e. a sequence of vectors) is denoted with
the bold style: $\boldsymbol{x}=(x_{1},x_{2},\ldots)$, where $x_{t}\in\mathbb{R}^{n_{x}\times1}$
and $t=1,2,\ldots$ indicates the discrete time; $x_{i,t}$ is the
$i$th component of the signal $\boldsymbol{x}$ at time $t$.

A regressor, i.e. a vector that, at time $t$, contains $n$ present
and past values of a variable, is indicated with the bold style and
the time index: $\boldsymbol{x}_{t}=\left(x_{t},\ldots,x_{t-n+1}\right)$.

The $\ell_{p}$ norms of a vector $x=\left(x_{1},\ldots,x_{n_{x}}\right)$
are defined as
\[
\left\Vert x\right\Vert _{p}\doteq\begin{cases}
\left(\sum_{i=1}^{n_{x}}\left|x_{i}\right|^{p}\right)^{\frac{1}{p}}, & p<\infty,\\
\max_{i}\left|x_{i}\right|, & p=\infty.
\end{cases}
\]
The $\ell_{\infty}$ norm is also used to denote the absolute value
of a scalar: $\left\Vert x\right\Vert _{\infty}\equiv\left|x\right|$
for $x\in\mathbb{R}$. 

The $\ell_{p}$ norms of a signal $\boldsymbol{x}=(x_{1},x_{2},\ldots)$
are defined as
\[
\left\Vert \boldsymbol{x}\right\Vert _{p}\doteq\begin{cases}
\left(\sum_{t=1}^{\infty}\sum_{i=1}^{n_{x}}\left|x_{i,t}\right|^{p}\right)^{\frac{1}{p}}, & p<\infty,\\
\max_{i,t}\left|x_{i,t}\right|, & p=\infty,
\end{cases}
\]
where $x_{i,t}$ is the $i$th component of the signal $\boldsymbol{x}$
at time $t$. These norms give rise to the well-known $\ell_{p}$
Banach spaces.

\section{NIC control design}

\label{sub:nl_des}

The proposed approach relies on identifying from the data \eqref{eq:data}
a model of the form
\begin{equation}
\begin{array}[t]{l}
\hat{y}_{t+1}=f\left(\boldsymbol{y}_{t},\boldsymbol{u}_{t}\right)\equiv f\left(\boldsymbol{q}_{t},u_{t}\right)\\
\boldsymbol{q}_{t}=\left(y_{t},\ldots,y_{t-n+1},u_{t-1},\ldots,u_{t-n+1}\right)
\end{array}\label{eq:model}
\end{equation}
where $u_{t}$ and $y_{t}$ are the system input and output, and $\hat{y}_{t}$
is the model output. For simplicity, the model is supposed of the
same order as the system but this choice is not necessary: all the
results presented in the paper hold also when the model and system
orders are different. Indications on the choice of the model order
are given in Section \ref{sec:design}.

A parametric structure is taken for the function $f$:
\begin{equation}
f\left(\boldsymbol{q}_{t},u_{t}\right)=\sum_{i=1}^{N}\alpha_{i}\phi_{i}\left(\boldsymbol{q}_{t},u_{t}\right)\label{eq:bfe}
\end{equation}
where $\phi_{i}$ are basis functions and $\alpha_{i}$ are parameters
to be identified. The basis function choice is in general a crucial
step, \cite{LLju1,HsNoAUT06,Novara2011711}. In the present NIC approach,
polynomial functions are used. The motivations are mainly two: (1)
polynomials have been proved to be effective approximators in a huge
number of problems; (2) as we will see later, they allow a ``fast''
controller evaluation. The identification of the parameter vector
$\alpha\doteq(\alpha_{1},\ldots,\alpha_{N})$ can be performed by
means of convex optimization, as shown in Section \ref{sec:design}.

Once a model of the form \eqref{eq:model} has been identified, the
controller $K^{nl}$ is obtained by its inversion:

Suppose that, at a time $t>0$, the reference value for the time $t+1$
is $r_{t+1}$ and the current regressor is $\boldsymbol{q}_{t}$.
Inversion consists in finding a command input $u_{t}^{nl}$ such that
the model output at time $t+1$ is ``close'' to $r_{t+1}$: 
\begin{equation}
\hat{y}_{t+1}=f\left(\boldsymbol{q}_{t},u_{t}^{nl}\right)\cong r_{t+1}.\label{eq:inv1}
\end{equation}
The latter equality may be not exact for two reasons: (1) no $u_{t}^{nl}\in U$
may exist for which $\hat{y}_{t+1}$ is exactly equal to $r_{t+1}$;
(2) values of $u_{t}^{nl}$ with a limited $\ell_{2}$ norm may be
of interest, in order to have a not too high command activity. This
kind of inversion is called (approximate) right-inversion and can
be performed also when $f$ is not bijective with respect to $u_{t}$
(e.g., for some $r_{t+1}$ and $\boldsymbol{q}_{t}$, more than one
value of $u_{t}$ may exist such that \eqref{eq:inv1} holds).

The command input $u_{t}^{nl}$ yielding \eqref{eq:inv1} can be computed
according to the following optimality criterion: 
\begin{equation}
\begin{array}[t]{ccl}
u_{t}^{nl} & = & \arg\min_{\mathfrak{u}\in U}J\left(\mathfrak{u}\right)\\
 &  & \textrm{subject to}\;\;\mathfrak{u}\in U.
\end{array}\label{eq:opt2}
\end{equation}
The objective function is given by
\begin{equation}
J\left(\mathfrak{u}\right)=\frac{1}{\rho_{y}}\left(r_{t+1}-f\left(\boldsymbol{q}_{t},\mathfrak{u}\right)\right)^{2}+\frac{\mu}{\rho_{u}}\mathfrak{u}^{2}\label{eq:objf2}
\end{equation}
where $\rho_{y}\doteq\left\Vert \left(\tilde{y}_{1-L},\ldots,\tilde{y}_{0}\right)\right\Vert _{2}^{2}$
and $\rho_{u}\doteq\left\Vert \left(\tilde{u}_{1-L},\ldots,\tilde{u}_{0}\right)\right\Vert _{2}^{2}$
are normalization constants computed from the data set \eqref{eq:data},
and $\mu\geq0$ is a design parameter, allowing us to determine the
trade-off between tracking precision and command activity. Indications
on the choice of $\mu$ are given in Section \ref{sec:design}.

Note that the objective function \eqref{eq:objf2} is in general non-convex.
Moreover, the optimization problem \eqref{eq:opt2} has to be solved
on-line, and this may require a long time compared to the sampling
time used in the application of interest. In order to overcome these
two relevant problems, a technique is now proposed, allowing a very
efficient computation of the optimal command input $u_{t}^{nl}$.

Since a polynomial basis function expansion has been considered for
$f$, it follows that the objective function $J\left(\mathfrak{u}\right)$
is a polynomial in $\mathfrak{u}$. The minima of $J\left(\mathfrak{u}\right)$
can thus be found considering the roots of its derivative: Define
the set
\[
U^{s}\doteq\left(\textrm{Rroots}\left(\frac{dJ\left(\mathfrak{u}\right)}{d\mathfrak{u}}\right)\cap U\right)\cup\left\{ \underline{u},\overline{u}\right\} 
\]
where $\textrm{Rroots}\left(\cdot\right)$ denotes the set of all
real roots of $\cdot$, and $\underline{u}$ and $\overline{u}$ are
the boundaries of $U$. The optimal command input is given by
\begin{equation}
u_{t}^{nl}=K^{nl}\left(r_{t+1},\boldsymbol{q}_{t}\right)\doteq\arg\min_{\mathfrak{u}\in U^{s}}J\left(\mathfrak{u}\right)\label{eq:opt3}
\end{equation}
where it has been considered that $U^{s}$ depends on the reference
$r_{t+1}$ and regressor $\boldsymbol{q}_{t}$. 

The nonlinear controller $K^{nl}$ is fully defined by the control
law \eqref{eq:opt3}. \medskip{}

\begin{remark}\label{rem:comp}The derivative $dJ\left(\mathfrak{u}\right)/d\mathfrak{u}$
can be computed analytically. Moreover, $U^{s}$ is composed by a
``small'' number of elements: 
\[
\textrm{card}\left(U^{s}\right)<\deg\left(J\left(\mathfrak{u}\right)\right)+2
\]
where card is the set cardinality and deg indicates the polynomial
degree. The evaluation of $u_{t}^{nl}$ through \eqref{eq:opt3} is
thus extremely fast, since it just requires to find the real roots
of a polynomial whose analytical expression is known and to compute
the objective function for a ``small'' number of values. This fact
allows a very efficient controller implementation on real-time devices.$\qquad\blacksquare$\end{remark}\medskip{}

\begin{remark}The system \eqref{ss_sys} is not required to be stable
and in general no preliminary stabilizing controllers are needed.
The only guideline is to generate the data using input signals for
which the system output does not diverge. This can be easily done
for many nonlinear systems like the single-corner model considered
below. Indeed, many systems are characterized by trajectories that
are unstable but bounded (a typical feature of chaotic systems). In
the presence of unbounded trajectories, for which a suitable input
signal can hardly be found, a preliminary stabilizing controller may
be required. The preliminary controller can also be a human operator,
who is able to drive the system within a bounded domain, see \cite{NoFaMiAUT13,movie_dfk}.$\qquad\blacksquare$\end{remark}

\section{Model identification algorithm}

\label{sec:design}

In this section, the algorithm for identifying the model \eqref{eq:model},
required by the nonlinear controller $K^{nl}$, is proposed. Systematic
criteria for the choice of the involved identification/design parameters
are also provided.

Choose a set of polynomial basis functions $\phi_{i}$. For example,
these functions can be generated as products of univariate polynomials,
where the independent variables are scaled to range in the interval
$[-1,1]$. In most cases, no large polynomial degrees are required:
we observed in several simulated and real-world applications that
a degree $\lesssim8$ is sufficient to guarantee satisfactory model
accuracy and control performance.

Define 
\[
\begin{array}[t]{c}
\tilde{\boldsymbol{y}}\doteq(\tilde{y}_{t_{1}+1},\ldots,\tilde{y}_{t_{2}+1})\\
\\
\Phi\doteq\left[\begin{array}{ccc}
\phi_{1}\left(\tilde{\boldsymbol{y}}_{t_{1}},\tilde{\boldsymbol{u}}_{t_{1}}\right) & \cdots & \phi_{N}\left(\tilde{\boldsymbol{y}}_{t_{1}},\tilde{\boldsymbol{u}}_{t_{1}}\right)\\
\vdots & \ddots & \vdots\\
\phi_{1}\left(\tilde{\boldsymbol{y}}_{t_{2}},\tilde{\boldsymbol{u}}_{t_{2}}\right) & \cdots & \phi_{N}\left(\tilde{\boldsymbol{y}}_{t_{2}},\tilde{\boldsymbol{u}}_{t_{2}}\right)
\end{array}\right]
\end{array}
\]
where $t_{1}\doteq1-L+n$, $t_{2}\doteq-1$, and $\tilde{u}_{t}$
and $\tilde{y}_{t}$ are the input-output measurements of the data
set \eqref{eq:data}. Consider the set $SC\subset\mathbb{R}^{N}$,
defined as
\[
\begin{array}[t]{r}
SC(\gamma_{y},\eta,\rho)\doteq\{\beta:\left\vert \widetilde{y}_{l+1}-\widetilde{y}_{k+1}+\left(\mathbf{\Phi}_{k}-\mathbf{\Phi}_{l}\right)\beta\right\vert \\
\leq\gamma_{y}\rho\left\Vert \widetilde{\boldsymbol{y}}_{l}-\widetilde{\boldsymbol{y}}_{k}\right\Vert _{\infty}+2\eta\rho,k\in\mathcal{T},l\in\Upsilon_{k}\}
\end{array}
\]
where $\mathcal{T}\doteq\{t_{1},\ldots,t_{2}\}$ and $\Upsilon_{k}$
is the set of indexes given by 
\[
\Upsilon_{k}\doteq\left\{ i:\left\Vert \tilde{\boldsymbol{u}}_{k}-\tilde{\boldsymbol{u}}_{i}\right\Vert _{\infty}\leq\zeta\right\} 
\]
and $\zeta$ is the minimum value for which every set $\Upsilon_{k}$
contains at least two elements. $SC$ is defined by a set of linear
inequalities in $\beta$ and is thus convex in $\beta$.

The parameter vector $\alpha\doteq(\alpha_{1},\ldots,\alpha_{N})$
of the model defined by \eqref{eq:model} and \eqref{eq:bfe} can
be identified by means of the following algorithm, completely based
on convex optimization. Note that the algorithm is ``self-tuning'',
in the sense that all the required parameters are chosen by the algorithm
itself, without requiring extensive trial and error procedures.\medskip{}

\textbf{Algorithm 1}\medskip{}

Initialization: choose a ``low'' model order (e.g. $n=1$).\medskip{}

\begin{enumerate}
\item \label{est_delta}Construct the vector $\tilde{\boldsymbol{y}}$ and
matrix $\Phi$ as indicated above.\medskip{}

\item \label{enu:eta}Compute
\[
\eta=\min\limits _{\beta\in\mathbb{R}^{N}}\left\Vert \widetilde{\mathbf{y}}-\mathbf{\Phi}\beta\right\Vert _{\infty}.
\]

\item \label{st_1}Solve the optimization problem
\begin{equation}
\begin{array}{l}
\alpha=\arg\min\limits _{\beta\in\mathbb{R}^{N}}\left\Vert \beta\right\Vert _{1}\\
\text{subject to}\\
(a)\ \beta\in SC(\gamma_{y},\eta,\rho)\\
(b)\ \left\Vert \widetilde{\mathbf{y}}-\mathbf{\Phi}\beta\right\Vert _{\infty}\leq\eta\rho
\end{array}\label{opt21a}
\end{equation}
where $\mathbf{\Phi}_{k}$ denotes the $k$th row of the matrix $\mathbf{\Phi}$,
$\gamma_{y}$ is the minimum value for which the constraints $(a)$
and $(b)$ are feasible, and $\rho$ is a real number slightly larger
than $1$.\medskip{}

\item \label{enu:mod_ord}Repeat steps \ref{est_delta}-\ref{st_1} for
increasing model order. Stop when no significant reductions of $\gamma_{y}$
are observed.\medskip{}

\item \label{enu:rho}Repeat steps \ref{st_1}-\ref{enu:mod_ord} for increasing
$\rho$. Stop when $\gamma_{y}<1$.$\qquad\blacksquare$\medskip{}

\end{enumerate}
The algorithm allows the achievement of three important features:\medskip{}

\begin{enumerate}
\item \textbf{\emph{Closed-loop stability}}. As proven in \cite{NoFaMiAUT13},
under reasonable conditions, constraint $(a)$ ensures that the function
$\Delta\doteq g-f$ has a Lipschitz constant wrt $\boldsymbol{y_{t}}$
non larger than $\gamma_{y}$, as $L\rightarrow\infty$. On the other
hand, a theoretical analysis shows that having this constant smaller
than $1$ is a key condition for closed-loop stability. Another required
condition is that $\Gamma_{y}<1-\gamma_{y}$, where $\Gamma_{y}$
is the input-output gain of the system formed by the cascade connection
of the controller and the model (this latter working in prediction).
Since $\Gamma_{y}$ can be imposed arbitrarily (see the discussion
below), it can be concluded that \emph{Algorithm 1 is able to ensure
closed-loop stability when the number of data becomes large}.\medskip{}

\item \textbf{\emph{``Small'' tracking error}}. Constraint $(b)$ is aimed
at providing a model with a ``small'' prediction error (this error,
evaluated on the design data set, is given by $\left\Vert \widetilde{\mathbf{y}}-\mathbf{\Phi}\alpha\right\Vert _{\infty}$).
According to the mentioned theoretical analysis, reducing this error
allows us to obtain a ``small'' tracking error. Note that there is
a trade-off between stability and tracking performance: In step \ref{enu:rho},
$\rho$ is increased until the stability condition is met. However,
increasing $\rho$ causes an increase of the prediction error and,
consequently, of the tracking error.\medskip{}

\item \textbf{\emph{Model sparsity}}. In step \ref{st_1}, the $\ell_{1}$
norm of the coefficient vector $\beta$ is minimized, leading to a
sparse coefficient vector $\alpha$, i.e. a vector with a ``small\textquotedblright{}
number of non-zero elements, \cite{Fuchs05,Donoho06_2,Candes06_1,Tropp06}.
Sparsity is important to ensure a low complexity and a high regularity
of the model, limiting at the same time well known issues such as
over-fitting and the curse of dimensionality. Sparsity allows also
an efficient implementation of the model/controller on real-time processors,
which may have limited memory and computation capacities.
\end{enumerate}
\hfill{}$\blacksquare$\medskip{}

Once a model has been identified, the nonlinear controller $K^{nl}$
is obtained by its inversion, as explained in Section \ref{sub:nl_des}.
Only one design parameter needs to be chosen for this inversion: the
weight $\mu$ in \eqref{eq:objf2}. If no particular requirements
on the activity of the command input $u_{t}$ have to be satisfied,
the simplest choice is $\mu=0$. Otherwise, if the input activity
has to be reduced, a value $0<\mu\leq\bar{\mu}$ can be chosen, where
$\bar{\mu}$ is the maximum value for which the stability condition
$\Gamma_{y}<1-\gamma_{y}$ holds. $\Gamma_{y}$ is the input-output
gain of the system formed by the cascade connection of the controller
and the model. This condition can be checked (approximately) by deriving
an estimate $\hat{\Gamma}_{y}$ of $\Gamma_{y}$ from the data \eqref{eq:data}.
Let
\begin{equation}
\mathcal{D}^{\Gamma}\doteq\left\{ \tilde{w}_{t},\hat{y}_{t+1}\right\} _{t=1-L+m}^{-1}\label{eq:data_gamma}
\end{equation}
where 
\begin{equation}
\begin{array}[t]{l}
\hat{y}_{t}=f\left(\tilde{\boldsymbol{y}}_{t-1},\tilde{\boldsymbol{u}}_{t-1}^{nl}\right)\\
\tilde{u}_{t-1}^{nl}=K^{nl}\left(\tilde{y}_{t},\tilde{\boldsymbol{q}}_{t-1}\right)\\
\tilde{\boldsymbol{q}}_{t-1}=\left(\tilde{y}_{t-1},\ldots,\tilde{y}_{t-n},\tilde{u}_{t-2}^{nl},\ldots,\tilde{u}_{t-n}^{nl}\right)\\
\tilde{w}_{t}=\left(\tilde{y}_{t},\ldots,\tilde{y}_{t-m+1}\right),
\end{array}\label{eq:stiga}
\end{equation}
$\tilde{u}_{t}$ and $\tilde{y}_{t}$ are the input-output measurements
of the data set \eqref{eq:data}, and $m\gg n$. The estimate $\hat{\Gamma}_{y}$
can be obtained applying the validation method of \cite{MiNoAUT04}
to the data set \eqref{eq:data_gamma} (the method is summarized in
the Appendix). Observing that $\tilde{u}_{t-1}^{nl}\equiv\tilde{u}_{t-1}^{nl}\left(\mu\right)$
and thus $\hat{\Gamma}_{y}\equiv\hat{\Gamma}_{y}\left(\mu\right)$,
$\mu$ must be chosen in such a way that $\hat{\Gamma}_{y}\left(\mu\right)<1-\gamma_{y}$.\medskip{}

\begin{remark}The stability conditions $\gamma_{y}<1$ and $\Gamma_{y}<1-\gamma_{y}$
can give indications on the choice of the control system sampling
time $T_{s}$: As discussed in \cite{Goodwin2010}, a too small $T_{s}$
leads to models where $\hat{y}_{t+1}\cong y_{t}$. These kinds of
models have a strong dependence on past outputs and a weak dependence
on the input, resulting in large values of $\gamma_{y}$ and $\Gamma_{y}$.
It is thus expected that $\gamma_{y}$ and $\Gamma_{y}$ can be reduced
by increasing $T_{s}$. Clearly, to capture the relevant dynamics
of the system and allow a prompt control action, $T_{s}$ must be
not ``too large''.$\qquad\blacksquare$\end{remark}

\section{Appendix: Nonlinear Set Membership validation procedure}

\label{appendix1}

In this appendix, the validation method of \cite{MiNoAUT04} is summarized,
suitably adapted for the present setting. This method is useful within
the NIC approach for estimating the constant $\Gamma_{y}$ appearing
in Sections \ref{sec:design} and \ref{sec:ibc_approach}. 

Suppose that $\Gamma_{y}$ is the Lipschitz constant of an unknown
function $\mathfrak{f}$. Let a set of data $(\widetilde{w}_{t},\hat{y}_{t+1})$,
$t\in\mathcal{T}$ be available, described by
\[
\hat{y}_{t+1}=\mathfrak{f}\left(\widetilde{w}_{t}\right)+d_{t},\quad t\in\mathcal{T}
\]
where $\mathcal{T}$ is a suitable set of indexes and $d_{t}$ is
a noise. This noise may also include errors due to the fact that $\mathfrak{f}$
is not Lipschitz continuous (e.g. in the case where $\mathfrak{f}$
is the sum of a Lipschitz continuous function plus a discontinuous
but bounded function). Assume that $d_{t}\in B_{\varepsilon}$, where
$B_{\varepsilon}$ is the $\ell_{\infty}$ ball with radius $\varepsilon$,
and that $\mathfrak{f}\in\mathcal{F}\left(\Gamma_{y}\right)$, where
$\mathcal{F}\left(\Gamma_{y}\right)$ is the set of Lipschitz continuous
functions on the domain of $\widetilde{w}_{t}$ with constant $\Gamma_{y}$.
Under this assumption, we have that $\mathfrak{f}\in FFS$, where
$FFS$ is the Feasible Function Set.\medskip{}

\begin{definition} \emph{Feasible Function Set}:
\[
\begin{array}{r}
FFS\doteq\{f\in\mathcal{F}\left(\Gamma_{y}\right):\hat{y}_{t+1}-f\left(\widetilde{w}_{t}\right)\in B_{\varepsilon},\ t\in\mathcal{T}\}.\\
\qquad\blacksquare
\end{array}
\]
 \end{definition}\medskip{}

According to this definition, $FFS$ is the set of all functions consistent
with prior assumptions and data. As typical in any identification/estimation
theory, the problem of checking the validity of prior assumptions
arises. The only thing that can be actually done is to check if prior
assumptions are invalidated by the data, evaluating if no function
exists consistent with data and assumptions, i.e. if $FFS$ is empty.\medskip{}

\begin{definition} \label{Def:validation}Prior assumptions are validated
if $FFS\neq\emptyset$.$\qquad\blacksquare$ \end{definition}\medskip{}

The following result provides necessary and sufficient conditions
for prior assumption validation. Define the function $\overline{f}\left(\Gamma_{y},w\right)\doteq\min\limits _{t\in\mathcal{T}}\left(\hat{y}_{t+1}+\varepsilon+\Gamma_{y}\left\Vert w-\widetilde{w}_{t}\right\Vert \right).$\medskip{}

\begin{theorem} \label{fals}A sufficient condition for prior assumptions
\\
to be validated is: 
\[
\overline{f}\left(\Gamma_{y},\widetilde{w}_{t}\right)>\hat{y}_{t+1}-\varepsilon,\ t\in\mathcal{T}.
\]
 \end{theorem}\medskip{}

\textbf{Proof.} See Theorem 1 in \cite{MiNoAUT04}.$\qquad\blacksquare$\medskip{}

The validation Theorem \ref{fals} can be used for assessing the value
of the Lipschitz constant $\Gamma_{y}$ so that the sufficient condition
holds. Suppose that $\varepsilon$ has been chosen by means of any
criterion (e.g. based on some prior knowledge on the noises, or by
means of standard filtering/smoothing techniques, or also using the
dispersion function defined in \cite{MiNoCDC05}). The constant
\begin{equation}
\Gamma_{y}^{\min}\doteq\inf_{\overline{f}\left(\Gamma,\widetilde{w}_{t}\right)>\hat{y}_{t+1}-\varepsilon,\ t\in\mathcal{T}}\Gamma\label{eq:gamin}
\end{equation}
represents the minimum Lipschitz constant for which the prior assumptions
are validated. A reasonable estimate of $\Gamma_{y}$ is thus a value
slightly larger than $\Gamma_{y}^{\min}$. Note that the evaluation
of $\Gamma_{y}^{\min}$ is quite simple, as shown by the examples
in \cite{MiNoAUT04} and \cite{MiNoTAC05}.

\bibliographystyle{IEEEtran}
\bibliography{lettnos_journals,lettnos_conferences,lettaltr,sparsification}

\end{document}